# Automated Tracking of Primate Behavior

Benjamin Y. Hayden[1]*, Hyun Soo Park[2], Jan Zimmermann[1]

**Affiliations**
1. Department of Neuroscience, Center for Magnetic Resonance Research, Center for Neuroengineering
2. Department of Computer Science and Engineering,
University of Minnesota, Minneapolis MN 55455

**\*Corresponding author**
Benjamin Yost Hayden
Department of Neuroscience
Center for Magnetic Resonance Research
University of Minnesota
Minneapolis, MN, 55455
benhayden@gmail.com

**Acknowledgements**
This work was supported by an award from MNFutures to HSP and BYH, from the Digital Technologies Initiative to HSP, JZ, and BYH, , by an R01 from NIMH (MH125377) to BYH, by am NSF CAREER (1846031) to HSP and by a P30 from NIDA (P30DA048742) to BYH and JZ.

**ABSTRACT**

Understanding primate behavior is a mission-critical goal of both biology and biomedicine. Despite the importance of behavior, our ability to rigorously quantify it has heretofore been limited to low-information measures like preference, looking time, and reaction time, or to non-scaleable measures like ethograms. However, recent technological advances have led to a major revolution in behavioral measurement. Specifically, digital video cameras and automated pose tracking software can provide detailed measures of full body position (i.e., pose) of multiple primates over time (i.e., behavior) with high spatial and temporal resolution. Pose-tracking technology in turn can be used to detect behavioral states, such as eating, sleeping, and mating. The availability of such data has in turn spurred developments in data analysis techniques. Together, these changes are poised to lead to major advances in scientific fields that rely on behavioral as a dependent variable. In this review, we situate the tracking revolution in the history of the study of behavior, argue for investment in and development of analytical and research techniques that can profit from the advent of the era of *big behavior*, and propose that zoos will have a central role to play in this era.

**Introduction**

The Minnesota Zoo in Apple Valley, MN, has the largest public collection of snow monkeys (*Macaca fuscata*) in the United States (**Figure 1**). Every morning, its twenty-seven macaques emerge from their dormitory and enter a large, beautifully architectured open enclosure. There are many ways to describe what they do next, but one way to say it is that they proceed to generate an *enormous* amount of data. That is, each of the monkeys moves each of its limbs in a specific way, moves its body position along a particular and often complex path, and interacts with multiple items in the pen. They also interact with each other in complex ways, they play, explore, foraging, relax, eat, and so on.

The ability to track and analyze the actions of these monkeys - and others - has potential relevance to researchers in biology and biomedicine, as well as for neuroscience, psychology, and in comparative biology (e.g., Rudebeck et al., 2019; Buffalo et al., 2019; Bliss-Moreau and Rudebeck, 2020; Santos and Rosati, 2015; Periera et al., 2020). And yet, nearly all the rich data they generate largely slips past us without being registered. Instead, studies of macaque behavior are limited either to what humans can annotate in hand-crafted ethograms or, if we want something more rigorous, to the monkeys' interactions with specialized response systems, such as levers and buttons. Such systems produce data that is several orders of magnitude lower in information rate than their full behavioral repertoire. For example, a measure of looking gaze direction or preference, typical in many psychological studies, gives one bit of information (left vs. right) per trial, which typically means a few hundred bits per hour. Those data can, of course, be used to test important hypotheses. But they obscure information beyond simple preference, such as motivation, arousal, attentional locus, and vigor (e.g., Niv et al., 2007; Shadmehr et al., 2019). In contrast, behavioral tracking can produce high quantities of data (for example, thirteen keypoints in 3D sampled at 30 Hz, Bala et al., 2020) without human intervention. Indeed, without tracking, our ability to collect behaviors with high technical precision and reproducibility is so limited that it might be described as looking at the world through a drinking straw.

This is all changing, and at a rapid pace. A recent series of technological advances has made it possible to collect a good deal of the data these monkeys produce. And not just these monkeys - other species and other locations can also be tracked (Dankert et al., 2009; Mathis et al., 2020; Joska et al., 2021; Walter and Couzin, 2021; Bala et al., 2020; Marshall et al., 2021).

We propose that the change is so profound that our new era needs a special name - we call it the era of ***big behavior***.

By our naming convention, little behavior would be measures of behavior that involve highly reduced, low-information tracking of what the animal is doing. Little behavior would include reaction times, preferences, gaze directions, even pupil size - measures that give a small amount of information per unit time. Little behavior would also include ethograms made by trained annotators. Big behavior ideally includes the entire body or much of it, e.g., full body pose (a set of landmarks specifying body joint locations) and involves continuous movement over an extended period of time. Scientists can now use multiple high-resolution cameras to continuously capture every fine-grained behavior over several weeks, months, and years, including both individuals and groups. As with any big data situation, the big behavior revolution raises unprecedented challenges, especially in managing, analyzing, and understanding the data in a fully automatic fashion. However, the benefits it offers are so great that many teams are working to solve them.

The present review will describe how we got to this point, talk about the state of the art, make a few predictions about the near future, and sketch out some of the potential benefits. We will discuss some of the specific scientific problems that big behavior for primates is likely to affect. And we will argue that zoos, often neglected in biomedical research, will likely take on a much larger and more central role in biological research in this new era.

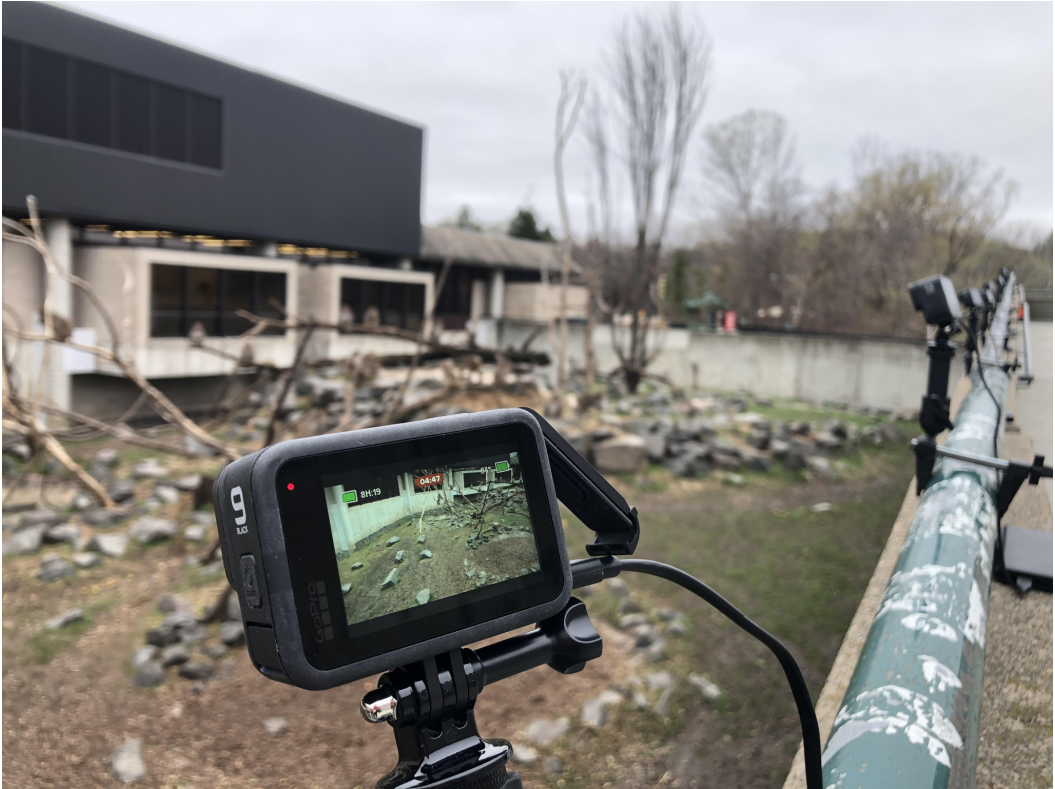

**Figure 1.** The Minnesota Zoo has a collection of 27 snow monkeys whose behavior provides a great deal of information about their internal states. That information is of great interest to biology and biomedicine. Relatively affordable cameras, such as GoPro cameras, when combined with computer vision tracking software, can provide excellent estimates of highly accurate pose.

**The challenges of big behavior in primates**

Behavioral tracking in primates has recently become possible through the parallel developments in several technological advances in computer vision, machine learning, and robotics. The story starts at the beginning of the last decade, when technical breakthroughs in deep learning enabled recognizing objects in an image using convolutional neural networks (CNNs, Krizhevsky et al, 2012). Subsequent works innovated on the design of CNNs to enable automatic tracking of humans both live and from videos (Cao et al., 2019 Newell et al., 2016; Wei et al., 2016; Fang et al., 2017). Parallel work allowed for tracking of animals such as flies, mice, and horses from videos (Mathis et al., 2018; Pereira et al., 2019; Mathis et al, 2020). The tools that allowed these animals to be tracked leveraged the copious capacity of CNNs to learn the visual and geometric relationship between landmark locations (body joints). They also relied on relatively cheap and robust digital cameras and standard computer vision techniques.

    Relative to other animals, non-human primates have been much more difficult to track (Bala et al., 2020; Negrete et al., 2021; Labuguen et al., 2020; Testard et al., 2021). There are three reasons for this. First, their body joints are highly flexible and thus generate a large number of distinctive body postures of which size is an order of magnitude larger than that of other species. Each joint has multiple degrees of freedom that constitute fine-grained activities such as bipedal/quadrupedal locomotion, hanging, and dexterous object manipulation. This makes a sharp contrast with non-primate animals such as rodents and insects - their poses can be characterized by a small basis set. Second, the body is covered by a textureless skin: it is a characteristic failure case of computer vision-based tracking algorithms. For example, the pelvis location is highly ambiguous while performing a sedentary activity because there are no visually or geometrically salient features to identify the pelvis joint. Third, their bodily movement is fundamentally three dimensional. The range of motion of bodies spans full three-dimensional space, and more importantly, the motion often involves 3D interactions with objects, peers, and environments. This induces considerable occlusion and diverse shape and appearances.

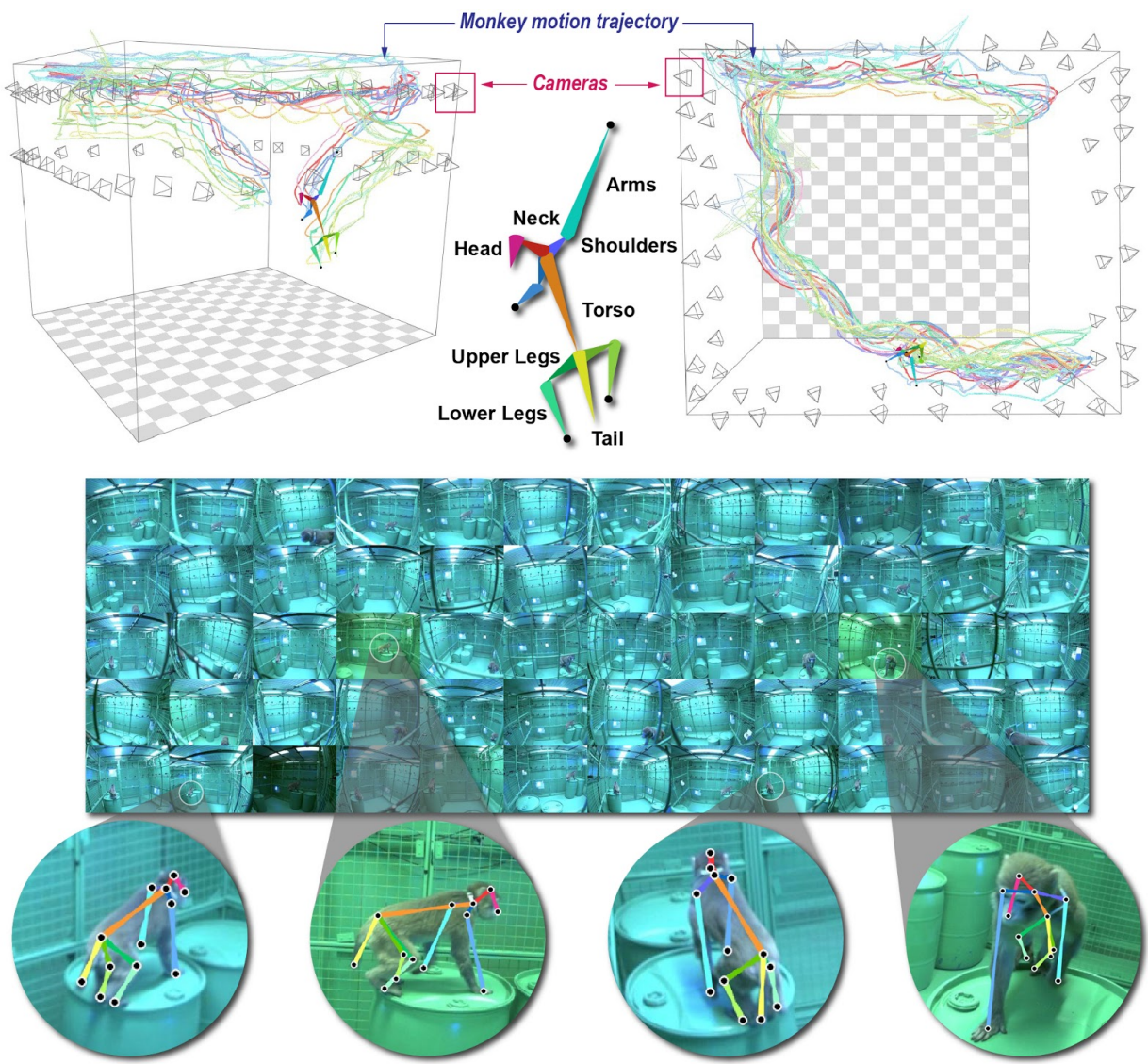

**Figure 2.** Automated pose-tracking software, such as OpenMonkeyStudio (Bala et al., 2020), can provide high quality tracking of poses in primates.

We have proposed that, given these difficulties, non-human primate tracking cannot be achieved using existing tracking paradigms. We need innovations in both hardware and software that are tailored to reflect the primate characteristics. For the readers' benefit, we next describe these briefly.

**(1) 2D vs. 3D representation**. Monkeys move in three dimensions. Actually, all animals move in three dimensions, but monkeys tend to make use of all three quite a bit - compare this with, say, a rodent, whose major axes of movement are on a plane along the ground. This is especially true in the laboratory, where the rat has a stereotyped behavior and little scientifically relevant movement in the z-dimension. With computer vision solutions, most behaviors have been represented in the two-dimensional (2D) camera/image coordinate system, i.e., a time series of 2D (XY) coordinates of behaviors. Continuing with the rodent example, a rodent is seen by an overhead camera where its XY location is tracked with respect to the image coordinate system. This 2D representation, however, shows limited expressibility to describe the behaviors of primates due to their 3D body movement over 3D scenes. The 2D representation is a camera projection of 3D behaviors where its location drastically varies as the viewpoint changes. A 3D representation (XYZ) is a viable solution that requires a specialized system such as a depth camera or a multiview camera system. For instance, a system of multiview cameras is used to develop OpenMonkeyStudio that enables tracking 3D body movements of macaques (Bala et al., 2020). The resulting 3D representation is invariant to the viewpoint change, which allows modeling coherent behavioral clusters compared with the 2D representation as shown in **Figure 3**.

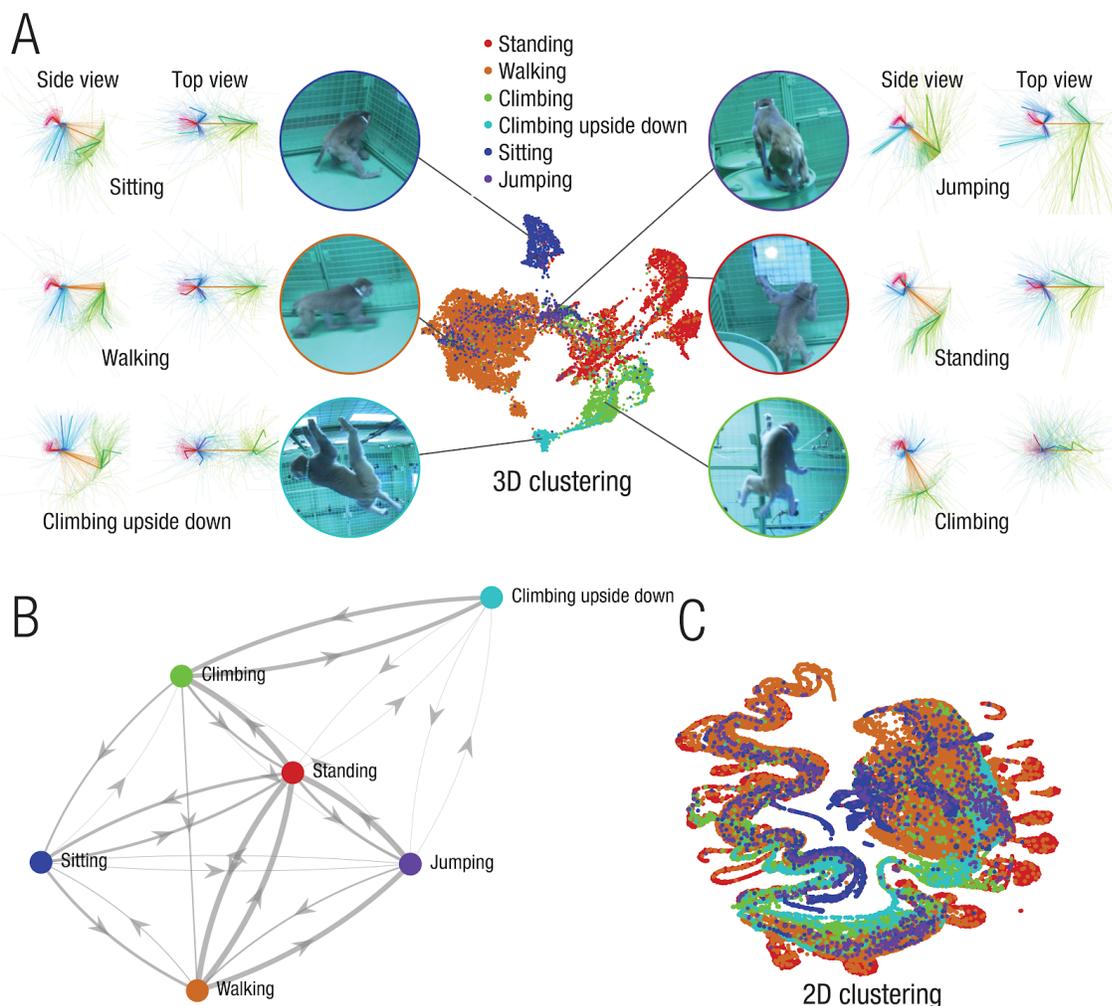

**Figure 3.** With OpenMonkeyStudio, we compare 2D and 3D representation by its ability to recognize semantic actions (standing, walking, climbing, climbing supine, sitting, and jumping). A. The poses are clustered by using UMAP. Each cluster that is represented by 3D poses (side and top views) is highly correlated with the semantic actions. B. With the clusters, we recognize actions in a new testing sequence using the k nearest neighbor search and visualize the transitions among the semantic actions. C. In contrast, the 2D representation provides the clusters that are driven by the pose and viewpoint. For instance, while the 3D representation of walking is one continuous cluster, the 2D representation is broken apart into discrete groupings of repeated poses at different spatial locations.

**(2) Target-specific model vs. generalizable model.** The ability of a CNN model to track primates reflects our ability to train the model. And our ability to train is limited by the size and quality of the training set. Thus, the success of tracking algorithms is predicated on the existence of a large, accurate, and well-curated dataset. For instance, large scale annotated datasets such as

10COCO and MPII have been used to train CNNs to reliably detect human poses (Lin et al., 2014; Andriluka et al., 2014). However, such large datasets do not exist for most animal species.

This data challenge has been addressed by training a target-specific model facilitated by annotation tools. For example, user interfaces provided by DeepLabCut and LEAP have enabled temporally extending pose annotations by annotating a few examples (10-1000) in a video (Mathis et al., 2018; Pereira et al., 2019). These methods, which make use of transfer learning or its variants, work for various organisms like flies, worms, and mice (e.g., Mathis et al., 2018; Nath et al., 2019).

Despite its remarkable performance, this approach is not suited to track the behaviors of non-human primates because visual appearance significantly varies as a function of viewpoint, pose, and species (Bala et al., 2020). As a result, a CNN trained using a dataset from one viewpoint does not generalize to that from another viewpoint. Instead of learning such target-specific models, we have proposed a new paradigm that aims to learn a generalizable model by collecting a large, curated dataset. This curated dataset must include diverse images across species, background, illuminations, poses, viewpoints, and interactions, where the size of data is comparable to the human data size (order of millions). Once the large dataset is collected, a generalizable CNN can be trained to detect primate poses from out-of-sample videos regardless of camera configuration, background, and species. For example, we have been collecting more than 100,000 primate images with 17 landmark annotations from images of primates in natural habitats. The data are obtained from the internet, national primate research centers, and from zoos, and include 26 species (20 monkeys and 6 apes) as shown in Figure 4. This is, by far, the largest collection for non-human primates. With this dataset, we recently opened a new benchmark challenge called OpenMonkeyChallenge (http://openmonkeychallenge.com/) that facilitates an annual competition to develop a generalizable pose detection model. We believe that this dataset, and others like it, will be crucial for developing widely usable models for tracking non-human primates.

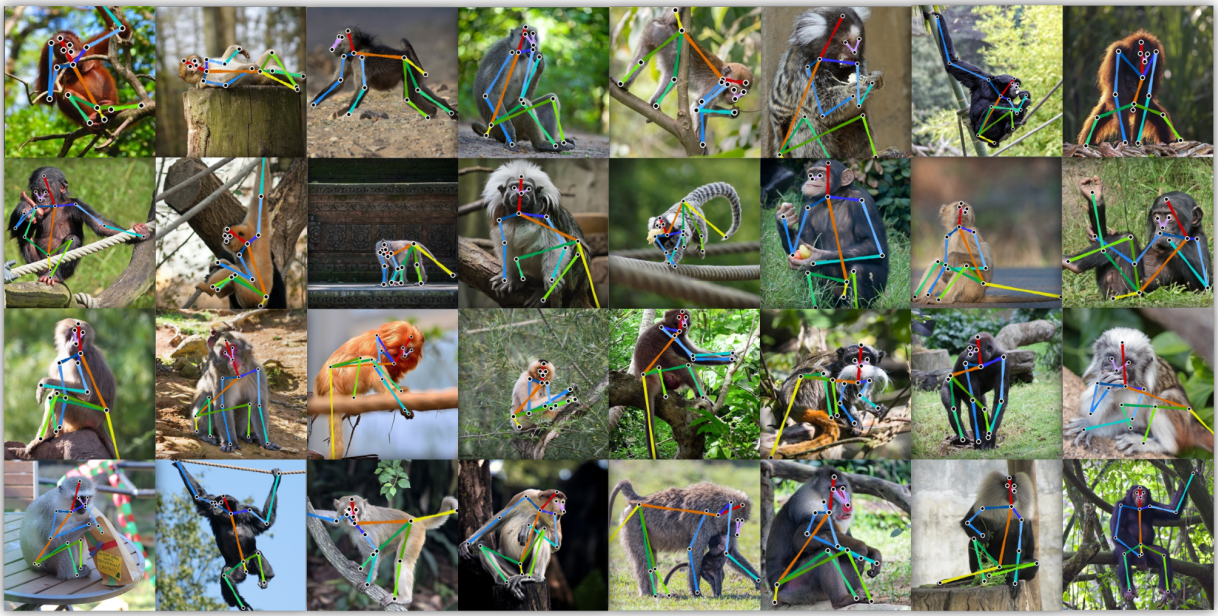

**Figure 4.** We argue that learning a generalizable model will facilitate big behavior for primates through a large collection of primate data across diverse species, pose, background, viewpoints, and illuminations. We collect a dataset called OpenMonkeyChallenge dataset to learn such a model via annual competition.

**(3) Single camera vs. multi-view camera system**: Existing animal tracking approaches are designed for an image stream from a monocular camera (ibid.). Due to the 3D nature of primate behaviors, a multiview camera system provides several benefits. These include

- *Disocclusion*: in tracking, occlusion is the main source of pose detection error. In particular, when a primate interacts with objects and other primates, by definition, it introduces inevitable occlusion. Indeed, one side of the animal's body will almost always occlude the other. This occlusion can be mitigated by a multiview camera system that offers observing the common primate from different viewpoints.
- *Robust estimation*: Multiple observations reinforce robust estimation. The multiview images of a common primate are visually and geometrically distinctive, and therefore, aggregating the detections from these multiview images allows more accurate and robust estimation in the presence of measurement noise.
- *Data augmentation*: Multiview images provide trivial data augmentation. Annotations from two viewpoints can be reconstructed in 3D, and in turn, projected onto other images, which forms additional annotations without manual effort. This leads to significant improvement on annotation efficiency (Marshall et al., 2021; Bala et al., 2020). For instance, in OpenMonkeyStudio, two to three images are manually annotated and nearly 50-60 multiview images are automatically annotated using the multiview geometry. This results in a generalizable model trained by a large annotated data (~200K images).

- 3D representation: Multiview geometry offers a 3D representation. This is different from the 3D lifting approach that reconstructs 3D pose from a single view image of which reconstruction is defined up to scale and orientation (Günel et al., 2019 lifting). With multiview images, full metric scale reconstruction can be obtained.

**What big behavior offers biologists**

The benefits of big behavior are difficult to predict but we will make some educated guesses. In any case, the promise is high (Datta et al., 2019; Krakauer et al., 2017; Calhoun and El Hady, 2021). Specifically, we envision that longitudinal and quantifiable analysis from years long recordings of primates will enable us to discover the reproducible behavioral patterns and characteristics, which has been impossible from a small scale data analysis. Here we walk through some of the potential benefits.

*Computational ethograms:* When scientists do want to quantify complex behavior, they have mostly used the ethogram approach. Behavioral imaging addresses the key limitations in this ethogramming approach by providing a computational solution that is cost-effective, consistent, fast, accurate, and fine-grained (Anderson and Perona, 2014).

(1) Cost-effectiveness: the ethogram approach involves highly trained human annotators who watch live animals or videos and mark down the behaviors they observe based on a predetermined rubric. These humans are skilled in identifying behaviors; indeed, the need to carefully train these observers is a major cost of the ethogram approach and one that makes it very expensive to implement. Being able to do that takes education and training and does not scale with large datasets. Big behavior automates behavioral annotations, which reduces the cost of training, management, and labor.

(2) Consistency and speed: different raters tend to have different subjective criteria, meaning the studies are often not reproducible with the same data if the raters are no longer available (Anderson and Perona, 2014; Levitis et al., 2009). Even well-trained individual human observers can often be somewhat inconsistent. This is especially likely to be a problem with novel and unusual behaviors, or ones that are poorly understood.

(3) Accuracy: human raters are intrinsically fallible - they get bored, especially after many hours of video footage, they use criteria that change subtly over time, they have blind spots, systematic patterns of behavior that they are less likely to detect. Consider for example how often referees make mistakes, or even disagree, in sports, where fair and foul are much more clearly defined. Humans are especially poor at detecting rare behaviors, and at detecting the kinds of behaviors that involve gradual change over long periods of time (Biggs et al., 2014; Kryszczuk and Boyce, 2002; Wolfe et al., 2005).

(4) Annotation resolution: humans have a sensitivity limit. We can only detect a certain number of behaviors and can only detect a certain granularity of behavior. For example, it may take some effort to be able to identify the different gaits of a running horse, especially subtly different ones. Moreover, it's possible that there are some patterns in the data that are real and measurable, but humans simply are not adapted to notice them without help. For example, some behaviors may reflect certain patterns of simultaneous activity across multiple limbs and may only be detectable and classifiable after certain dimensionality reduction processes that humans lack the cognitive capacity to perform.

Automated ethogramming procedures offer the promise of using a computer to perform the ethogramming procedure, rapidly, cheaply, and on a massive scale. Because it is much cheaper and faster, automated ethogramming promises many things that are uneconomic with human annotators, such as massive throughput testing, or monitoring behavior 24 hours/day, or testing unlikely hypotheses that would be too costly to test with standard approaches.

*Replacement for other dependent variables:* A good deal of research may require some measure of efficacy of some intervention, but not depend on any particular measure. For example, consider a team of researchers who want to rapidly screen several dozen candidates for headache medicine, most of which will be inert. They may be able to identify promising ones as ones that have some effect at all on, and further investigate those. In that case, behavioral imaging may provide information sufficient to move on to the second stage. And it may do so at a price that is much cheaper, and with less intervention, than other measures (such as measure or internal physiological functioning). Or for example, a researcher may want to know whether an animal differentiates seeing itself in the mirror from seeing another conspecific.

Indeed, in such cases, big behavior offers the possibility of performing such experiments in animals that are inaccessible to other measures, including rare and endangered animals, and highly intelligent animals, such as apes. Even for standard laboratory primates, big behavior may be an order of magnitude cheaper than other measures (see above) and may therefore be a preferred alternative. For example it may require less specialized equipment, trained technicians, or built laboratory environments. Finally, even if it is not cheaper, it may have fewer welfare costs than invasive measures, and may be preferred for that reason.

*Natural behaviors*: Finally, tracking allows scholars to monitor naturalistic - or at least relatively naturalistic behaviors. When primates perform tasks that resemble those for which they have evolved, their behavior is more likely to reflect their normal response biases, and thus to be more ethologically valid, and likely more interpretable (Pearson et al., 2014; Hayden, 2018). For example, we have argued that both risk and impulsivity measures in primates are biased by standard laboratory tasks (Hayden, 2016; Blanchard et al., 2013; Eisenreich et al., 2019).

**What big behavior offers medicine**

*Improvements in diagnosis:* Diagnosis often requires specialized expertise in humans; in animals, who cannot talk and often disguise their symptoms, it's often more difficult. It often occurs only following conspicuous presentation of behavioral symptoms often requires advanced veterinary intuition. In many cases, animals deliberately camouflage their symptoms. For these reasons, diagnosis is likely to be a major focus for big behavior. This is not to say that computers will monitor all behavior and replace veterinary expertise - the disappointing progress of diagnosis of x-rays serves as a cautionary tale. Instead, we imagine that big behavior will provide a complementary measure that will boost trained medical opinion. First, it will offer the ability to identify subtle and hidden behavioral patterns. For instance, a monkey may slightly drop her shoulder while walking when her arm is wounded. Such differences represent symptoms, which have not been coded and listed as a behavioral marker by the experts. With big behavior, it is possible to precisely characterize such subtle and hidden patterns through longitudinal observations. Second, it will offer acute measures, with greater quantity, accuracy, and sensitivity than human observers can provide. Third, it will provide diachronic measures, such as changes taking place over hours, days, or weeks, which can be difficult or expensive to collect. These benefits are especially likely to accrue in the case of large facilities with many individuals, such as at primate centers or pet sanctuaries.

*Validating animal models of disease:* Primate models of disease are crucial to modern biomedical research. In many cases, models are imperfect or are only relevant to a subset of symptoms. In other cases, the validity of primate models is unknown. For example, the limitations of models are especially well delineated in the case of psychiatric illness, which often relies on personal reports. Consider that many of the major criteria for depression and obsessive-compulsive disorder rely on subjective descriptions of feelings (Goodman et al., 1989; Beck et al., 1988). Ascertaining the validity of an animal disease model is often a surprisingly ad hoc procedure, based on superficial assessment of major symptoms (Geyer and Markou, 1995; Koob and Zimmer, 2012). In many cases, establishing a valid animal model for some disease or symptom class takes many years - even decades - to result in a consensus across the field. Despite these problems, any disease is characterized by a specific set of behaviors and behavioral patterns; it is possible that the particular constellation of behaviors associated with any disease can serve as an ethological fingerprint of the disease. Optimistically, these fingerprints can be compared across species to test specific hypotheses about the behavioral validity of specific models.

*Improvements in treatment*: Many diseases have well defined treatments that require parameter-setting (Johnson et al., 2013). For example Parkinson's Disease is well treated by deep brain stimulation, although the parameter space is very large. The optimization of parameters takes place through what is more or less a gradient descent procedure, that is, through trial and error (Schiff, 2010). The slow part is the assessment of state in the patient. Primate models can improve that process, but they still have a slow assessment stage - often even slower in primate because they lack the verbal modality. Big behavior offers a solution to this problem - it is both high bandwidth and computerized, so it can be made very fast, and potentially improve treatment time by orders of magnitude. It is possible - albeit speculative at this point - that behavioral imaging can lead to rapid closed loop parameters for treatment. We can test specific parameters and read out the effects of those parameters - whether they are stimulation parameters in DBS, anatomical positioning of stimulation, different drug doses, or any other of a large number of possibilities.

*Improvements in diagnosis*: Many diseases are really clusters if distinct conditions with somewhat different symptom profiles and different best treatments. For example, clinical depression is usually responsive to the third pharmaceutical treatment tried; each try takes several months, so if diagnosis could be sped up, a great deal of suffering could be eliminated. The ability to characterize behavior with high bandwidth promises greater information that can be clustered and specific disease subtypes identified. Even modest improvements in classification of diseases into different subtypes, which in turn can motivate specialized study of treatment. To give one well-known example, deep-brain stimulation (DBS) treatment for depression in humans is sometimes focused on the subgenual anterior cingulate sulcus (Mayberg et al., 2005; Lozano et al., 2008). This is a brain region whose responses are known to be associated with depression, with negative outcome monitoring, and with depression severity (Alexander et al., 2019; Roberts and Clarke, 2019; Drevets et al., 2008; Azab and Hayden, 2018). However, despite promising early treatment results, follow-up studies have been equivocal. It has been proposed that the core issue is that we are treating two different diseases with different anatomical bases (Drysdale et al., 2017; McGrath et al., 2013). Improved diagnosis in humans would help us test these hypotheses faster, and more valid animal models (see above) would help even more.

**The role of zoos in the era of big behavior of primates**

Most primate research in biology and biomedicine takes place in the laboratory. In contrast, zoos have historically been an underutilized resource. Zoos do have some appeal for research. They have many captive primates, these are well studied and well characterized, and are around for a while. Indeed, many zoos are eager to collaborate with scientists. The underutilization of zoos stems from several specific factors that, on balance, give the laboratory distinct advantages. These include the freedom to perform invasive experiments and access to animals full time, the ability to use a large, in principle unlimited, number of animals, and the ability to focus on a small number of well-studied research organisms, with the benefits that standardization of research animals brings. With the advent of big behavior, however, several of the features - including weaknesses - of zoos are poised to become strengths. As such, zoos are likely to become more important in research. Here, we cover several specific features of zoodeeps that make them likely to be of increased importance in the era of big behavior:

*Zoos are designed for natural behavior:* Zoos are designed with public audiences in mind, and are architectured to maximize unobtrusive viewing from many vantage points. Indeed, they are laid out to have viewing in a way that does not bother animals and is also safe from their interference. This is ideal for big behavior. It allows measuring natural behaviors including social interactions with minimal interference (see above). This is an advantage over many research laboratory environments, which are often constructed generically and without observation in mind. Many laboratories have tight space restrictions, which can result in imaging that is difficult or distorted. This is also a major advantage of zoos relative to field research sites, which pose many practical problems for observation. For example, field sites are often remote and far from power (and studies can last multiple weeks), and repair facilities.

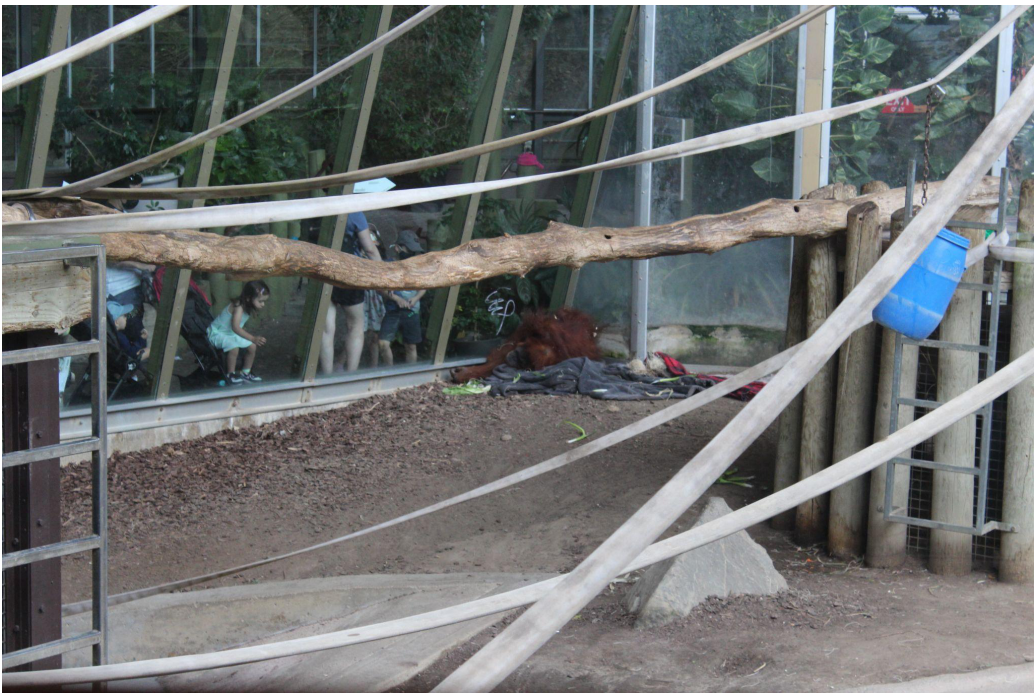

**Figure 4.** The orangutan enclosure at the Toronto Zoo allows for panoramic viewing of the research subjects, who have a large and enriched space to move in. Such enclosures are particularly useful for behavioral imaging.

*Zoos have many individuals*: While laboratory studies of primates typically focus on few individuals (in many cases, two per study), zoos have access to many more. This number is augmented by the national network of exchange of animals between zoos. Of course, laboratories can compensate for low numbers with a high degree of training and instrumentation per subject. But with tracking, these needs can be reduced. The larger number of individuals is important because it can reduce Type II errors, can improve generalizability, and can facilitate individual difference studies. Moreover, the relative ease of collecting behavioral imaging data means that data from multiple zoos can in principle be combined, thereby further increasing numbers. This can be done with low cost because zoos are supported by attendees and patrons, and so the marginal cost of adding additional animals is low. In contrast, laboratory research is supported solely by funding for the research itself, so the cost of adding additional animals to a study is borne by the research funds.

*Zoos have many species*: Individual laboratories specialize in a handful of model organisms and focus on deep search into the biology of those individuals. In contrast, zoos typically have a large number of species, the better to interest the public. This breadth of species has several benefits. Perhaps the greatest is for comparative studies. Indeed, the lack of variety of species has been identified as a limitation of major biological and biomedical research in the modern era, and a limit on the generalizability of the resulting science. The variety is limited for practical reasons by the complexity of having to learn the intricacies of multiple animal species. Zoos already bear this cost. We can only imagine the benefits to research of having access to a large number of species, each of which can be selected for a particular study tailored to the specific needs of that study.

*Zoos have excellent records about their animals*: Zoos already have access to their animals for long periods of time, which makes it possible to answer important questions about their ancestry, behavioral history, their DNA, endocrine profile, and so on. This is an advantage relative to field studies, where animal subjects are often catch-as-catch-can. Even in cases where the same individuals are followed for many years, there is an enormous cost to developing this knowledge, and that cost makes it harder to explore new species or groups of individuals. The record-keeping in zoos is also better than the laboratory, where most animals are procured from suppliers with unknown details, or, in the case of primates from breeding sites where this information is often unavailable.

*Zoo environments are designed for welfare*: Zoos have a keen interest in keeping their animals happy and unstressed. While zoos are not perfect environments for their animals, they often do a great job with enrichment, variety, safety, and other factors that reduce stress and boredom, and keep animals at least somewhat energized and happy. Zoos have a strong incentive to do this- audiences like seeing happy animals, and like seeing animals move around and interact with their milieux. In contrast, laboratory environments are designed to optimize data collection. As a result, animals in zoos are more likely to be unstressed, and provide ethologically valid behavior. That type of behavior is invaluable. For example, psychiatric studies often rely on the assumption that animal models are psychiatrically normal - deviations introduce biases which can in turn reduce the efficacy of treatment effects. Even non-psychiatric studies, however, can be hampered by the introduction of stressors that serve as uncontrolled variables (at best) or as competitors to variables of interest.

**Conclusion**

The ability to track pose in animals, especially primates, is poised to provide many research benefits to biologists and medical researchers. The data such methods provide is greatly enriched compared to data derived from earlier methods, such as preference, reaction time, and gaze direction, which involve low amount of information about animals' internal states. This is not to minimize the value of such methods. Measure of low dimensional behavior has been critically important for a large number of studies. Indeed, that approach has been a mainstay of our own labs. Just to give a few examples in primates from our own lab, preference measures alone have given important information about risk preferences, impulsivity, memory, cognitive capacity, and so on - often when used in conjunction with complex models (Ebitz et al., 2019; Blanchard et al., 2014; Farashahi et al., 2019). Our labs, among others, have been interested in studying more complex and naturalistic behaviors and have begun to do so, although, but still using very simple measures (e.g. Cash-Padgett and Hayden, 2020; Eisenreich et al., 2019). Clearly big questions can be asked with conventional approaches.

However, it's still looking at the world through a keyhole. With new technologies and new approaches, we are now able to open the door and step into the world and see behavior fully. It's still hard to predict all the effects that change will cause. That's partly because most of the analyses to study these data have not been invented. They will require new mathematical techniques that are only beginning to be delineated. Moreover, it's not clear how much of the animals' internal states leak out through their behavior. Nonetheless, we believe that these changes will lead to great advances in our understanding of the machinery of living things.